# Blockchain Developments and Innovations – An Analytical Evaluation of Software Engineering Approaches

Mahdi Fahmideh[1], Anuradha Gunawardana[2], Shiping Chen[3], Jun Shen[2], and Brian Yecies[2]

[1]University of Southern Queensland, Australia
[2] University of Wollongong, Wollongong, Australia
[3]CSIRO Data61, Eveleigh, Australia
Mahdi.Fahmideh@usq.edu.au, anuradhawg66@gmail.com, Shiping.Chen@data61.csiro.au, jshen@uow.edu.au, byecies@uow.edu.au

**Abstract.** Blockchain has received expanding interest from various domains. Institutions, enterprises, governments, and agencies are interested in Blockchain's potential to augment their software systems. The unique requirements and characteristics of Blockchain platforms raise new challenges involving extensive enhancement to conventional software development processes to meet the needs of these domains. Software engineering approaches supporting Blockchain-oriented developments have been slow to materialize, despite proposals in the literature, and they have yet to be objectively analyzed. A critical appraisal of these innovations is crucial to identify their respective strengths and weaknesses. We present an analytical evaluation of several prominent Blockchain-oriented methods through a comprehensive, criteria-based evaluation framework. The results can be used for comparing, adapting, and developing a new generation of Blockchain-oriented software development processes and innovations.

## 1     Introduction

Based on recent trends and evidence, views on the development of Blockchain technology are diverse and contrasting. A recent global survey by Deloitte [1] found that IT-based organizations are keen to adopt Blockchain and consider adopting it as a priority to improve the performance of their operational systems. Furthermore, Bosu et al. [2] reported the results of a prototype Blockchain project hosted on GitHub that more than doubled in engagement from 3,000 to 6,800 use cases between March and October 2018. The number of related projects launched within this relatively short time demonstrates the allure of this technology among our research community.

On the other hand, substantial financial losses caused by numerous attacks and system failures related to Blockchain and smart contract applications are evident in many industry reports. Notable examples include the Coinrail exchange hack in 2018, with the loss of $42 million worth of cryptocurrencies; the DAO attack in 2016, ending in the withdrawal of Ether funds worth $50-60 million; the $65 million loss following the Bitfinex attack in 2016; and the $600 million loss due to the 2014 MtGox attack [3, 4]. To mitigate such failures, adopting systematic software engineering approaches, as acknowledged in several previous studies (e.g., [4, 5]), is essential. A systematic engineering methodology will allow Blockchain developers to design a Blockchain system and implement it in a manageable manner without exposing it to attacks and vulnerabilities. Moreover, unlike an ad-hoc methodology, errors occurring within a systematic approach can be better traced and fixed. A systematic approach will better assist development teams to deal with the uncertainties surrounding Blockchain-oriented software caused by its relative immaturity and the many under-explored areas associated with the technology.

Responding to these issues, in this paper we set up a research agenda to i) review existing advances in Blockchain development; ii) propose an evaluation framework including a coherent set of criteria derived from both the Blockchain and software engineering literature; iii) evaluate the selected development approaches against the criterion set; and iv) outline evaluation outcomes. Hence, our study contributes to Blockchain-oriented software engineering in two major ways:

- By providing an evaluation framework as a useful tool by which to compare and contrast existing Blockchain engineering approaches and to prioritize and select one innovation which fits the requirements of a given Blockchain-oriented system development project.
- By identifying unaddressed knowledge gaps in the innovations relating to Blockchain development in order to map out future research directions.

Section II explores the history and background of Blockchian technology and discusses recent work on software engineering for Blockchain-based systems. Section III presents a review of a selected set of Blockchain development innovations. Section IV details the criteria for an evaluation framework, along with an evaluation of existing Blockchian development approaches. In Section V, we discuss the evaluation outcomes reported in the previous section, as well as the limitations of the processes reviewed. Finally, conclusions and suggestions for future work are presented in Section VI.

## 2 Background

### 2.1 Blockchain

Blockchain technology originated with the introduction of Bitcoin cryptocurrency in 2008 [6]. Since then, industrial interest in Blockchain system development has expanded significantly, with companies exploring the potential of Blockchain-enabled Internet-based systems for the future [7]. Fundamentally, a Blockchain is a cryptographically linked chain of records or blocks, with each block containing a hash value of the previous block and one or more transaction logs with their timestamp [7]. These chains of blocks are stored on a distributed node network, allowing each participant node to retain a copy of the Blockchain. Participating nodes validate each new block by collectively agreeing if the new block can join the existing Blockchain. The process of reaching collective agreement is known as a consensus mechanism. After successful validation, a new block is added to the existing Blockchain. These validating and chaining procedures make these blocks suitable for storing sensitive financial transaction information, as users can rely on a secure exchange of information without needing an intermediary, potentially a less trustworthy mechanism [8].

The ability to create smart contracts is an important attribute of Blockchain technology. Smart contracts are database slots that store the necessary logic to create and validate transactions; these contracts allow users to read, update, and delete data stored in Blockchain systems [6]. These smart contracts can be implemented either via domain-specific languages like Solidity on Ethereum, or using general-purpose languages like Java and Go, which can be familiar to Blockchain developers. Moreover, smart contracts create a pathway for non-Blockchain software systems to integrate Blockchain technology, where the business logic, rules, and data specific to that system are coded into smart contracts which are then executed and deployed in decentralized ledgers. However, the meticulous design and robust development of smart contracts in Blockchain systems are essential to mitigate the effects of malicious attacks and exceptions caused by poorly designed or badly implemented platforms.

### 2.2    Development of Blockchain-based Systems

Blockchain-based software engineering is associated with a range of concepts and terminologies. A common understanding of these diverse notions and terms is essential to successful Blockchain system development. According to Porru et al [4], a Blockchain-based system is a novel software system that utilizes a Blockchain implementation in its components. Thus, innovations across various Blockchain developments can be viewed as an extension of traditional software development, with the need to incorporate features of a Blockchain system such as decentralized architecture, systematic block transaction recording, and data redundancy [6, 9].

As mentioned in Section I, Blockchain development should be based on systematic approaches, characterized by an endorsed collection of phases, activities, practices, tools, documenting, and user training [10], thereby providing clarity about how one should perform each activity prescribed under a given process. Although adopting such methodologies may not necessarily guarantee optimal software quality, as suggested in [11, 12] there is a strong correlation between the quality of a particular engineering innovation and the final software product's performance.

In developing Blockchain systems, developers encounter numerous challenges including, but not limited to, compromise between security and performance, choice of an appropriate consensus mechanism, and the complexities around multiple stakeholder corporations [7, 13]. On top of these challenges, the relative immaturity of Blockchain technology increases the complexity of Blockchain adoption, calling for extra effort from developers used to working on conventional software engineering projects. As pointed out by Ingalls [14] forty years ago, the more complex the system, the more susceptible it is to total breakdown, making it all the more important for developers to follow systematic engineering approaches incorporating the Software Development Life Cycle (SDLC). In this paper, the evaluation framework to be elaborated in Section IV has incorporated the complexities surrounding Blockchain adoption, and its criteria have been developed with a strong focus on the SDLC and recommended systematic software engineering practices.

## 3    Existing Studies of Blockchain-based Systems Development

This section briefly describes six prominent Blockchain development innovations, which have been selected based on four key criteria. Thus the approach: i) fully or partially describes the development process of a Blockchain system; ii) is based on all (or at least some) of the SDLC phas-es; iii) describes all Blockchain's chief integral characteristics discussed in Section II – for instance, smart contracts and block validation; iv) has been recently published, between January 2018 and December 2020. Based on our investigation of academic papers in line with these criteria, we selected six approaches, namely CBDG [15], BADAO [16], BSDP [17], BSCRE [19], BAFISCT [20], and BCSTM [21]. Since many studies are ongoing, this is an incomplete list. For each of these approaches, we provide a brief description of its development process, focusing on the SDLC phases.

### 3.1 CBDG

The CBDG approach [15] describes the development tasks required to build a Blockchain system, shown in Fig. 1. As the first task, possible future benefits of integrating Blockchain are identified. In this space, either the existing systems are migrated to a Blockchain-enabled system or a completely new system is developed from scratch. If integrating Blockchain is considered beneficial, the next task is to select a suitable Blockchain implementation platform like Ethereum. The authors of [15] underline the importance of using such a platform as against building a completely new Blockchain, which could potentially involve many years of work.

The third task involves identifying development requirements and defining an appropriate Blockchain model and a conceptual workflow. A range of other related factors – including i) permissions from the Blockchain network, ii) choice of front-end programming languages, iii) external databases, and iv) servers – are also considered. Next, a Proof-of-Concept (PoC) prototype of the Blockchain system is designed to secure client approval. In designing this PoC, client feedback is also incorporated. The formulated prototype consists of various components including i) information architecture, ii) designs, and iii) sketches. Once the PoC is approved by the client, visual and technical designs are completed as the fifth task. These artefacts depict the complete design of the Blockchain system to be developed, and they also incorporate User Interface and API designs. The sixth and final task entails developing the Blockchain system based on the set designs. Here the first development is referred to as the pre-alpha version, as formal testing and client approval have not yet been realized. The pre-alpha version is then subjected to thorough testing and moves through three more versions, alpha, beta, and finally the Release Candidate version. At the end of this process, the fully tested system is deployed. Importantly, the deployed Blockchain system should be able to be upgraded when required.

### 3.2 BADAO

The BADAO approach [16] describes a model-based, process-driven method for developing a

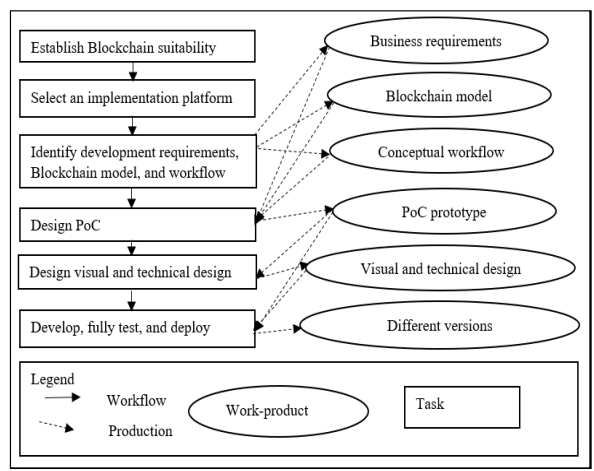

**Fig. 1.** CBDG approach block diagram

Blockchain-enabled system – either a Decentralized Autonomous Organization (DAO) or Blockchain-Augmented Organization (BAO). In a DAO, the traditional centralized transaction processing is decentralized and automated via smart contracts. In the case of BAOs, they are identified as organizations, and are augmented with Blockchain features such as immutability and traceability.

Fig. 2 illustrates the development tasks germane to the BADAO approach. Firstly, a Business Process Model (BPM) for the desired business scenario is defined. This BPM guides the subsequent development process based on SDLC phases. The next task involves establishing the suitability of Blockchain for the identified business case, expressed as either DAO-suitable, BAO-suitable, or not suitable. If BAO is found to be suitable, the process boundaries of the BAO are determined. Here, consideration is given to automating as many processes as possible utilizing Blockchain and smart contracts, while allowing non-Blockchain processes to complement Blockchain-enabled ones.

After completing these tasks, the construction of a Platform Independent Model (PIM) is undertaken. This model is independent of any features specific to a particular Blockchain platform. However, the PIM includes features such as smart contract architectures, Blockchain state definitions, and security models attached to the Blockchain model. Next, a Platform Specific Model (PSM) is constructed to incorporate elements relevant to the Blockchain platform selected. The realized PSM can be used to implement the Blockchain solution once the smart contracts are implemented and the design concepts are validated.

### 3.3 BSDP

BSDP [17] has undertaken an online survey of 1604 Blockchain developers in 145 Blockchain projects hosted on GitHub. The survey asked about the different methods utilized by Blockchain developers in conducting requirement analysis, tasks assignment, testing, and verification of Blockchain projects. These development tasks are depicted in Fig. 3.

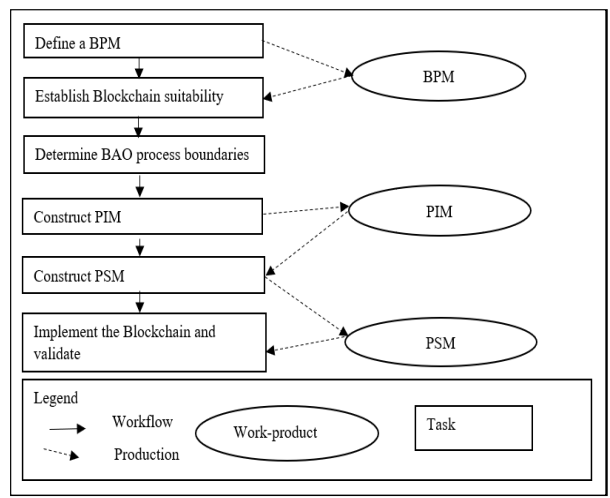

**Fig. 2.** BADAO approach block diagram

In terms of requirement analysis, most requirements are identified by project managers and through community discussion. In these discussions, ideas are brainstormed among community members via online and offline meetings. Customer feedback and the selection of requirements by developers themselves are other forms of requirement analysis. Regarding task assignment in Blockchain projects, few options are identified. Allowing developers to select tasks based on personal preference is one option. Some tasks are assigned based on developer expertise – BSDP points to the relative inexperience and unfamiliarity of developers in dealing with Blockchain projects.

Regarding testing, the BSDP survey revealed that unit testing and code review were the two main code quality assurance innovations utilized in Blockchain projects. Unit tests are either written by developers themselves or by a separate quality assessment team. Manual testing of the code by developers themselves is another popular testing mechanism identified. In addition, functional testing is utilized to test the functionality of the end software against established system requirements. Moreover, a separate Testnet, which is an alternative Blockchain, can be deployed to test the security and scalability of Blockchain projects without breaking the main Blockchain.

### 3.4 BSCRE

BSCRE describes the design and implementation of a Blockchain system in the real-estate industry [19]. A graphical overview of BSCRE's development tasks is provided in Fig. 4. Firstly, the requirements of the proposed Blockchain system are gathered. Next, the design of smart contracts involves three main steps: i) formulating actors and their role definitions, ii) defining business service functions, and iii) describing Ethereum processes.

Regarding actors and their roles, two main actors named as contract owner and users are identified. The contract owner is usually the real-estate owner who is responsible for the development of the smart contracts. Users or tenants create their own Ethereum wallets to access the Blockchain network. Turning to business services functions, smart contracts require four main functions: i) creation of new transactions, ii) generation of smart contracts, iii) sending messages, and iv) mining using Ethereum. Concerning the Ethereum processes, [19] identifies four: i) block validation,

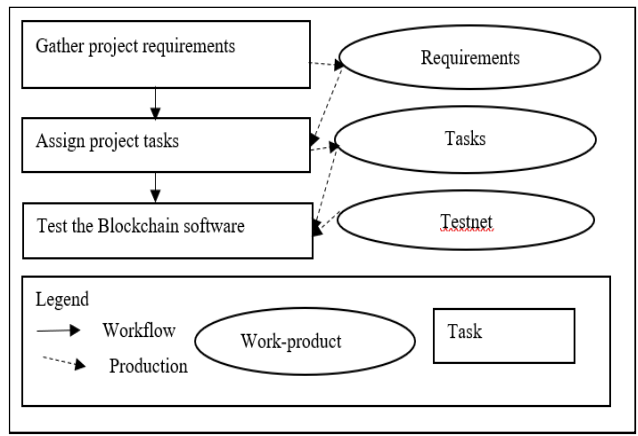

**Fig. 3.** BSDP approach block diagram

777

ii) network discovery, iii) transaction creation and iv) mining. All validated blocks join the peer-to-peer Blockchain network via the network discovery process. Further, the mining process ensures that all new validated blocks are added to the Blockchain and broadcast to the whole network.

After designing smart contracts as described above, they are implemented on a suitable Blockchain network like Ethereum. A dapp is also developed if the Blockchain system requires a User Interface. Once smart contracts are implemented, they are compiled to generate a binary file. Next, the contracts are deployed on an Ethereum network using Ethereum clients. Finally, a Web application is developed to interact with the smart contracts.

### 3.5 BAFISCT

BAFISCT describes a development process designed to integrate Blockchain with supply chain processes [20]. The tasks associated with BAFISCT's development process are shown in Fig. 5. Firstly, the target product for the supply chain operations is defined. This is followed by the identification of the characteristics of the selected product. These product characteristics include a range of factors – for instance, the producer, price, and design of the product. The third task entails identifying all the requirements attached to the product, which can be functional, regulatory, or technical. Based on these requirements, the main actors involved in supply chain processes relevant to the selected product are defined as the fourth task. Next, the different operations and processes attached to these actors are identified and modelled as *Block Flow Diagrams.*

Following this step, the business rules relevant to the product and its operations are defined. These rules are included in the Blockchain, and will be appropriately executed to process supply chain transactions relevant to the product. Next, the different digital assets relevant to supply chain processes are also defined. Following this, the information flow within the identified digital assets and processes are defined. Once the information flow is recognized, a complete view of a Blockchain transaction in terms of the information processed and its subsequent outcome on Blockchain can be observed.

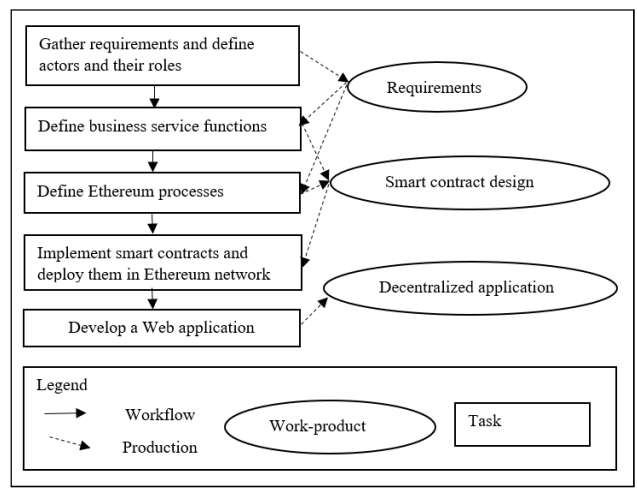

**Fig. 4.** BSCRE approach block diagram

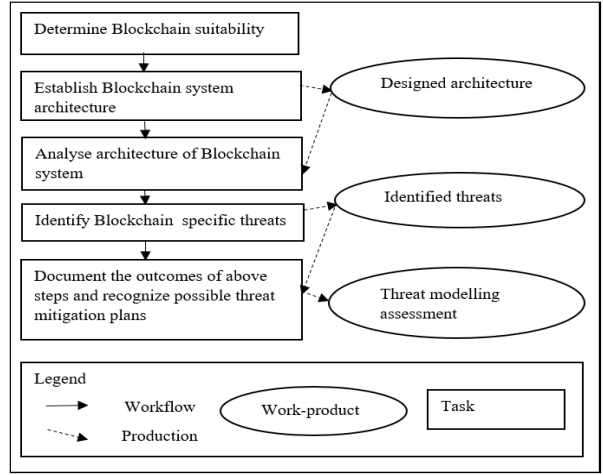

**Fig. 5.** BAFISCT approach block diagram

The next task, configuration of Blockchain, involves i) identification of a suitable Blockchain network (permissioned or permission-less), ii) selecting a suitable consensus mechanism and a Blockchain platform, iii) designing User Interfaces, and iv) developing APIs. Finally, testing of the configured Blockchain via unit and integration tests is performed.

### 3.6 BCSTM

The BCSTM approach introduced by [21] is designed to be used in conducting a security assessment of Blockchain-enabled software architecture. For that purpose, it identifies a range of Blockchain-specific security threats and, based on these threats, the selected architecture is evaluated utilizing the popular STRIDE threat-modelling approach [22]. Fig. 6 illustrates the development tasks associated with the BCSTM approach.

Firstly, [21] discusses a range of factors that impact the suitability of Blockchain for a given scenario. Here, among many other factors, Blockchain features such as immutability, and basic Blockchain functions such as block validation, are also considered. After establishing Blockchain suitability, the next task is to define a Blockchain architecture and select an appropriate Blockchain implementation. For implementation, a suitable network – for instance, a permissioned network – should be selected. Further, regarding data storage, [21] describes three possible options. These are *hash*, where only the hash value of a data item is stored on Blockchain; *generic*, for all non-hash data storage; and *smart contract*, for the storage of executable code.

Next, Blockchain-specific threats relevant to the selected Blockchain architecture are identified. In [21] eight separate categories have been identified to indicate the range of these threats; smart contract, cryptocurrency, and permissioned ledger threats are a few of the categories considered. Finally, based on these listed threats, a threat-modelling assessment is conducted to generate a holistic view of Blockchain security. In this assessment, possible threat mitigation actions and decisions are also recognized and documented.

## 4 Criteria-based Evaluation

### 4.1 Evaluation Framework

Our developed evaluation framework is structured to review existing Blockchain approaches, and to classify, evaluate, and characterize their innovations based on accepted software engineering practices. In so doing, we have followed two main steps as described below.

Step I. Defining meta-level characteristics: Meta-characteristics are features that are anticipated will be satisfied by an ideal evaluation framework. It is essential to have a set of meta-characteristics to guide the selection of appropriate criteria for the framework, as they can be used to evaluate different criteria and decide whether they should be added to the framework. For the purpose of our framework, we extracted five meta-characteristics defined in [23]. These characteristics are i) preciseness, for creating unambiguous, quantifiable, and descriptive criteria; ii) simplicity, for ease of understanding; iii) soundness, for the relation or semantic link between the criterion and the problem domain; iv) minimal overlapping, for distinct and minimally interdependent criteria; and v) generality, to ensure the abstract character of criteria independent of specific details, standards, and technologies.

Step II. Derivation of the criteria set: We reviewed existing evaluation frameworks such as [23, 25, 26], as well as more recent Blockchain literature, to derive a set of criteria which are applicable to Blockchain development and also satisfy the meta-characteristics defined in step I above. Following an iterative refinement and elimination of duplicated criteria, a list of eighteen criteria was derived. Table I briefly describes each of these eighteen criteria, which were utilized to evaluate all the approaches evaluated in Section III.

The criteria selected span eight categories. Four cover the 'analysis', 'design', 'testing and implementation', and 'deployment' phases of the SDLC. Two criteria, modelling language and work products, are associated with the 'modelling' category since they capture different representational languages and models applicable to Blockchain development. The 'user support and training' category includes criteria that provide support and guidance for developers to create a Blockchain system. Unsurprisingly, the tool criterion comes under the 'tool support' category. Finally, four additional criteria are classified under the 'other' category, exceptional features which address other elements anticipated in a Blockchain development approach. Although we are not suggesting that our framework covers all possible criteria relevant to Blockchain development, we believe that such a comprehensive framework is not found in the existing Blockchain literature.

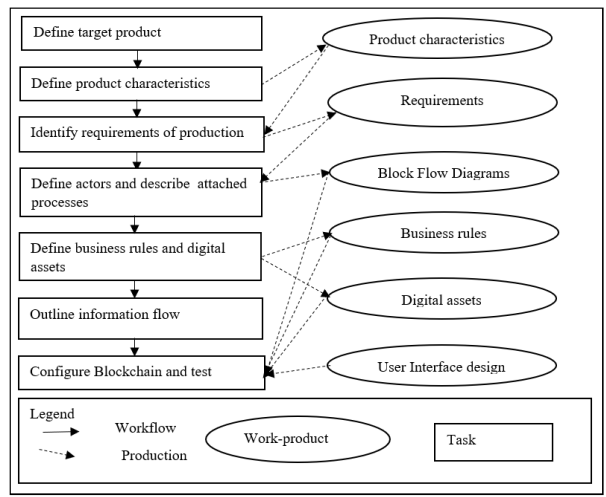

**Fig. 6.** BCSTM approach block diagram

**Table 1.** Evaluation Criteria

| Criteria description/evaluation questions (Letter C uniquely identifies the criteria) ||
|---|---|
| *Criteria related to the analysis phase* ||
| **Analysing context (C1):** Does the approach describe factors that are used to determine suitability of integrating Blockchain with a software system? | **Requirement analysis (C2):** Does the approach describe or refer to a requirement-gathering process, techniques, or methods? |
| *Criteria related to the design phase* ||
| **Smart contract design (C3):** Does the approach describe or refer to a smart contract design process? Is the functionality of a smart contract described? | **Consensus mechanism (C4):** Does the approach refer to a consensus protocol used and/or describe a functionality in a Blockchain system? |
| **Architecture design (C5):** Does the approach describe the overall architecture of a Blockchain system? Has the proposed architecture been segregated into multiple layers? | **Security (C6):** Is there any discussion of enhancing or maintaining security of a Blockchain system and architecture design requirements? |
| **Privacy (C7):** Is there any discussion of how the privacy of user data is protected, or are there references to privacy risks, guidelines or policies applicable to a Blockchain system? ||
| *Criteria related to the implementation and testing phase* ||
| **Testing (C8):** What is the nature of the support, in terms of techniques and recommendations, provided by the approach in testing functional and non-functional operations? ||
| *Criteria related to tool support* ||
| **Tools (C9):** Is there any evidence of third-party or custom-made tools that can be used to speed up or automate tasks being followed in development of a Blockchain system? ||
| *Criteria related to the deployment phase* ||
| **Deployment mechanism (C10):** Does the approach refer to deployment of a Blockchain system? Is there any evidence of configuration of hardware and/or software components that are needed for deployment? ||
| *Criteria related to modelling* ||
| **Modelling language (C11):** Has the approach included one or more representational languages used at design and/or run time of a Blockchain system? | **Work products (C12):** Is there any evidence of one or more interim project outputs/artefacts applicable to each SDLC development phase? |
| *Criteria related to user support and training* ||
| **Training (C13):** Is there any evidence of training manuals, user documentation, or other forms of support and guidance to develop a Blockchain system? | **Procedures and supportive techniques (C14):** Does the approach include algorithms or step-by-step guidance to follow or practice tasks required to develop a Blockchain system? Is there any evidence of supportive techniques or examples related to development tasks? |
| *Other criteria* ||

| | Scalability (C15): Does the approach describe techniques/factors that allow a Blockchain system to scale up to handle high volumes of transactions and data requests, or refer to scalability testing mechanisms? | Blockchain type (C16): Does the approach identify or suggest a suitable Blockchain network for a Blockchain system? |
|---|---|---|
| | Domain applicability (C17): Is the approach directed towards one or more industries or domains? | Development roles (C18): Does the approach define or describe different roles required to develop a Blockchain system? |

## 4.2 Evaluation Outcomes

In Table II, the evaluation outcomes of the six Blockchain approaches based on 15 scaled criteria are summarized. The scaled criteria are based on a five-point Likert Scale: fully supported, considerably supported, moderately supported, slightly supported, and not supported. Three remaining criteria, C16, C17, and C18, are descriptive in nature, as the answers to them are more open-ended. Hence, they are not evaluated based on the scale. For C16, the type of Blockchain network supported under each process is reviewed. For C17, the target domain of each approach is scrutinized. For C18, a distinct list of development roles applicable to Blockchain development are extracted from the selected innovations.

**Table 2.** Evaluation Outcomes

| Criteria | Approach | | | | | |
|---|---|---|---|---|---|---|
| | *CBDG* | *BADAO* | *BSDP* | *BSCRE* | *BAFISCT* | *BCSTM* |
| C1 | ● | ◐ | ○ | ◔ | ○ | ● |
| C2 | ◐ | ◔ | ● | ◐ | ◐ | ◔ |
| C3 | ◔ | ◐ | ◔ | ● | ○ | ◐ |
| C4 | ◐ | ◔ | ◔ | ◐ | ◕ | ● |
| C5 | ◐ | ◐ | ◔ | ● | ● | ◐ |
| C6 | ◕ | ◐ | ● | ◔ | ◔ | ● |
| C7 | ◔ | ◔ | ○ | ◔ | ◔ | ◕ |
| C8 | ● | ○ | ● | ○ | ● | ○ |
| C9 | ● | ◐ | ◕ | ◔ | ◐ | ○ |
| C10 | ● | ○ | ◔ | ◐ | ○ | ○ |
| C11 | ○ | ● | ○ | ◐ | ◐ | ◐ |
| C12 | ● | ◐ | ◐ | ◐ | ● | ◐ |
| C13 | ○ | ○ | ◐ | ○ | ◔ | ◔ |
| C14 | ◐ | ◐ | ● | ◐ | ◕ | ● |
| C15 | ○ | ○ | ● | ○ | ◔ | ◐ |
| ● | Fully supported | | | | | |

| Criteria | Approach | | | | | |
|---|---|---|---|---|---|---|
| | *CBDG* | *BADAO* | *BSDP* | *BSCRE* | *BAFISCT* | *BCSTM* |
|  | Considerably supported | | | | | |
|  | Moderately supported | | | | | |
|  | Slightly supported | | | | | |
|  | Not supported | | | | | |

## 5 Criteria-based Evaluation

### 5.1 Findings

In this section, we briefly discuss the findings related to each criterion of our evaluation framework. For the 15 scaled criteria, the discussion is based on the evaluation outcomes reported in Table II.

Analyzing context (C1) – Due to the complexity surrounding Blockchain-enabled software development, a range of innovations should provide clear guidance in establishing the suitability of Blockchain for a given software system. Only the BCSTM and CBDG approaches fully satisfy this criterion. BCSTM and CBDG review a range of factors to establish Blockchain suitability [15, 21] including i) the requirement to store users' personal information on Blockchain itself; ii) the need to update the rules of the software system; iii) rewarding or compensating participating nodes; iv) the number of nodes required to validate new blocks; and v) required transaction speed. Furthermore, BSDP and BAFISCT fail to provide any information about this criterion.

Requirement analysis (C2) – identifies the functional and non-functional requirements that need to be fulfilled by a Blockchain system. Further, approaches may provide descriptions of supporting techniques, such as interviews and workshops, which can be used to gather requirements. BSDP is the only method to fully satisfy this criterion. It describes the different techniques used to gather Blockchain project requirements based on the findings of a survey of Blockchain projects hosted on GitHub. These techniques are briefly summarized in Section III under the review of the BSDP approach. Additionally, BSCRE considerably supported this criterion, as BSCRE mentions conducting organizational workshops and gathering requirements from the different stakeholders of a company. Notably, none of the reviewed methods achieved a rating of not supported.

Smart contract design (C3) – is an integral part of a Blockchain system. If they are not meticulously designed, the whole Blockchain system is susceptible to malicious external attacks. Only BSCRE was able to fully satisfy this criterion. The main steps include i) redefining actors based on their direct interaction with the smart contracts; ii) defining smart contract decomposition; iii) defining message flows and data structure; iv) defining modifiers (special functions called before other functions) and internal functions; and v) defining tests and security assessment procedures [19]. BSCRE provides a comprehensive smart contract design process, which is discussed under the BSCRE approach segment in Section III. BAFISCT is the only method to provide an absence of details on smart contract design.

Consensus mechanism (C4) – ensures that new blocks are only added to the Blockchain network once majority nodes agree and verify them. Although the role of a consensus mechanism is referred to in five out of the six reviewed approaches, only BCSTM achieved a rating of fully supported. Accordingly, the role of a consensus protocol is more critical in a permission-less network, as anyone can participate in its transaction validation process. Also in BCSTM is the need to continuously provide adequate financial compensation for all nodes participating in block validation in a permission-less network. If nodes are not adequately compensated, the block validation process will not run at optimum efficiency, which could result in malicious attacks on the Blockchain system. The alternative consideration to this problem is to use a permissioned network where the number of participating nodes is controlled [21]. Furthermore, the BAFISCT approach states that the chosen consensus mechanism should be compatible with the Blockchain platform or framework, such as Ethereum, on top of which the Blockchain system is to be developed.

Architecture design (C5) – of a Blockchain system provides evidence of how each component of the system is positioned relative to the other components. Architecture can also be described according to multiple layers. As Table II shows, two approaches fully satisfied this criterion, while no single approach was rated as not supported. A brief description of the Blockchain architecture utilized in each approach is provided in Section III.

Security (C6) – Security is an important dimension associated with Blockchain systems. Of the reviewed innovations, only BSDP and BCSTM fully satisfy this criterion. As elaborated in Section III, BCSTM proposes a threat modelling process designed to conduct a security assessment of a Blockchain-enabled software architecture. The outcomes of this threat-modelling assessment provide valuable insights into the level of security evident in the architecture of a given Blockchain. Moreover, BSDP has discovered that most Blockchain projects incorporate popular code quality assurance mechanisms such as unit testing and code review to test the security of a Blockchain system. It also mentions that bug bounty, static program analysis, simulation, and external audit [17] are used in this regard.

Privacy (C7) – Privacy of user data stored on Blockchain is another important dimension of Blockchain systems. Processes should consider widely accepted standards, rules, and policies on user data privacy when designing Blockchain systems. However, the reviewed approaches provide minimal details about this criterion. The BCSTM approach, which achieved a rating of considerably supported, is the highest rated. According to BCSTM, users' personal information should not be stored on public elements of the Blockchain as it can violate their privacy rights. Furthermore, malfunctions and defects in smart contracts can expose private user data to unauthorized parties [21]. Notably, as Table II shows, two approaches fail to provide any details about Blockchain privacy.

Testing (C8) – The testing mechanism describes the techniques and recommendations provided by the methods to test functional and non-functional operations of a Blockchain system. While three out of the six reviewed approaches fully satisfy this criterion, three others did not provide any details on testing. The testing mechanism associated with each supporting process is briefly described in Section III.

Tools (C9) – External third-party tools or custom tools can be used to automate or speed up the tasks involved in developing a Blockchain system. Except for BCSTM, all the approaches provide evidence of tool support. However, only CBDG fully supported this criterion, describing a wide range of tools that can be used to automate different Blockchain development tasks. For instance, the Truffle Ethereum framework can be used in developing dapps, and can also serve as a testing framework. Furthermore, the Solium tool is used to format code written in Solidity, and fix security issues in the code.

Deployment mechanism (C10) – The deployment of Blockchain systems can become complex as it requires the configuration of both hardware and software components. Further, the system should be fully tested before being deployed to a production environment. The reviewed approaches provide minimal details on deploying Blockchain systems, with three achieving a rating of not supported. CBDG is the only fully supported approach. Among other elements, it states that the deployed system should be able to receive upgrades in accordance with business requirements. It also mentions various tools that can automate deployment-related tasks. For instance, Remix IDE is a tool that can be used to deploy smart contracts. BSCRE mentions deploying implemented smart contracts to an Ethereum network using Ethereum clients, Geth and PyEthApp [19]. BSDP also refers to the deployment of a fully tested system despite failing to provide detailed descriptions.

Modelling language (C11) – A modelling language can be used to represent different work products in a Blockchain development innovation in a structured manner. Apart from two approaches, as Table II shows, the selected processes have all utilized some form of modelling language. However, only BADAO fully supported this criterion.

Work products (C12) – Work products are the interim project deliverables that can be identified from a Blockchain development process. All the reviewed methods incorporated at least one work product, and no approach received a rating lower than moderately supported. In Section III, we have modelled the work products relevant to each approach in block diagrams as shown in Fig. 1-6.

Training (C13) – Procedures should provide training, in terms of training manuals, user documentation, and other forms of support and guidance necessary to develop a Blockchain system. To our knowledge, none of the six reviewed approaches provides comprehensive details of training. This might be a serious limitation that needs to be considered by potential practitioners and researchers in the future. Nevertheless, a few of the lines of action provide partial support for this criterion. For instance, BCSTM supports documenting evaluation outcomes of its threat modelling assessment.

Procedures and supportive techniques (C14) – Step-by-step guidelines or an appropriate algorithm might assist developers to better understand the various development tasks described in a Blockchain innovation. In addition, some helpful examples, or supportive techniques designed to undertake these tasks might also be provided. All six reviewed approaches, as Table II shows, provide some level of support for this criterion. However, only three approaches achieved the highest rating.

Scalability (C15) –The ability of a Blockchain system to handle large volumes of data and transactions is a sign of its high scalability. However, only the BSDP approach fully satisfied this criterion, as it discusses a range of relevant testing techniques, such as stress testing. Otherwise, while a few strategies refer to scalability issues in Blockchain systems, none provides any details of possible mechanisms to mitigate them. Four of the reviewed approaches failed to provide any details on scalability.

Blockchain type (C16) – Table III identifies the supporting Blockchain network types.

Domain applicability (C17)  Table IV identifies the applicable arena for each approach.

Development roles (C18) – describe the duties and responsibilities of different IT professionals participating in a Blockchain system. However, development role definitions are limited in existing approaches. Table V summarizes the identified development roles.

**Table 3.** Blockchain Type

| Approach | Blockchain Network Type | | |
| --- | --- | --- | --- |
| | *Permission-less* | *Permissioned* | *Not Stated* |
| CBDG | | | ✓ |
| BADAO | | ✓ | |
| BSDP | | | ✓ |
| BSCRE | | ✓ | |
| BAFISCT | | | ✓ |
| BCSTM | | ✓ | |

**Table 4.** Domain Applicability

| Approach | Domain |
| --- | --- |
| BAFISCT | Supply chain |
| BSCRE | Real estate |
| BCSTM, BSDP, CBDG, BADAO | Not stated or multiple domains |

**Table 5.** Development Roles

| Role | Referred approaches | Description |
| --- | --- | --- |
| Smart contract owner | BSCRE | Responsibilities to create, compile, and deploy smart contracts |
| Software engineer | BADAO | To perform software engineering roles in developing a Blockchain-oriented software |
| Blockchain developer | BCSTM, CBDG, BADAO, BSDP | To implement Blockchain design models and code smart contracts |
| Quality assurance | BSDP | Quality checking/ testing of Blockchain software |
| Project lead | BSDP | Overseeing a Blockchain project, and define project requirements when needed |

### 5.2 Limitations

Based on the level of support for the criteria set for our evaluation framework, we identified a number of limitations among existing Blockchain development approaches.

Firstly, previous studies have raised concerns regarding the lack of a comprehensive development methodology to guide the development of Blockchain-based systems. Based on evaluation outcomes reported in Table II and the individual analytical analysis in Section III above, existing Blockchain development approaches limit their focus to a few selected SDLC phases, and their descriptions of Blockchain adoption are generally below the level expected of a full-scale methodology.

Secondly, existing approaches and methods provide very low support for training. Due to this limitation, developers, especially those without experience, following these approaches may cause problems that might result in poorly developed Blockchain systems.

Similarly, there is minimal support for the deployment phase of the SDLC. Existing approaches show little interest in deploying a fully tested Blockchain system, despite deployment being a complex phase requiring proper guidance.

Further, the selected innovations failed to define many of the development roles applicable to Blockchain development, and we were only able to extract five roles (see Table V).

Last but not least, there is inadequate discussion about protecting the privacy of user data. Since global regulators consider user data privacy a priority, the approaches examined should have given more attention to this issue.

## 6 Conclusions and Future Work

This paper underlines the need for systematic engineering approaches and innovations to develop Blockchain systems. As a first attempt to fill this need, we presented a descriptive and comprehensive review of six existing Blockchain development approaches in the context of a proposed evaluation framework. Our results highlighted both the strengths and shortcoming of existing approaches; areas for further improvement include phases, activities, practices, tools, documenting, and user training [10]. Future Blockchain applications should incorporate these requirements into their development process so as to ensure both the security and quality of the target Blockchain-oriented software.

Given these findings, a clear research direction for future investigations is the development of a comprehensive Blockchain software engineering innovation that would draw on the strengths of existing approaches, while avoiding their weaknesses. This broad aim can be achieved by extracting method fragments from older processes found in [e.g. 26, 18], as well as from existing Blockchain development approaches, and amalgamating them to create a fully-fledged methodology. Once crafted, the newer and more innovative approach can be customized and improved to accommodate the requirements of different Blockchain systems. Despite the unlikelihood of a single standard or agreement being reached in the future, this effort calls for cooperative work from experts in different fields.